$$$$This is a LaTex file, followed by a tex file ``informII.tex'' as well
as 15 Postscript files.  Between each of these files, there is a series of
``$$$''.  To print this mess out, I suggest using an editor such as emacs
to make 16 files from this one file.  All files, except ``informII.tex''
can have arbitrary names.  Make sure to remove my comments... they are
surrounded by ``$$$''

Good Luck
Brian Schmidt, 60 Garden St. Cambridge, MA 02138
brian@cfanewton.harvard.edu
$$$$  Main Body $$$$

\documentstyle[11pt]{article}
\input imformII
\begin{document}
\vskip .1 in
\centerline {\bf Expanding Photospheres of Type II Supernovae and the}

\centerline {\bf Extragalactic Distance Scale$^{1}$}

\vskip .4 in
\centerline {\bf Brian P. Schmidt and Robert P. Kirshner$^2$}

\centerline {Harvard-Smithsonian Center for Astrophysics}
\centerline {60 Garden St.}
\centerline {Cambridge, MA 02138}
\vskip .4 in
\centerline {\bf Ronald G. Eastman}

\centerline {Harvard-Smithsonian Center for Astrophysics}
\centerline {and}
\centerline {Board of Studies in Astronomy and Astrophysics, Lick Observatory}
\centerline {University of California, Santa Cruz}
\centerline {Santa Cruz, California 95064}

\vfill
\centerline{To appear in the August 20, 1992 issue of
{\it The Astrophysical Journal}}
\noindent $^1$Based in part on observations made at the Multiple
Mirror Telescope, jointly operated by the Smithsonian Institution and
the University of Arizona.
\vskip .05 in
\noindent $^2$Guest observer at Kitt Peak National Observatory, operated by
AURA, Inc., under contract to the National Science Foundation.
\eject
\pagenumbering{arabic}
\pagestyle{plain}
\centerline {\bf Abstract}

\noindent We use the Expanding Photosphere Method to determine distances to 10
type II supernovae.
The effects of asymmetries, extinction, and flux dilution are explored.
 Using empirical evidence and
time-independent, spherical models which treat H and He in non-LTE, we show
that blackbody
corrections caused by flux dilution are small for type II supernovae in the
infrared, and
in the optical when their color
temperatures are less than 6000~K. The extinction to a type II-P supernova
can be estimated from its light curve: the uncertainty introduced into a
distance measurement due to extinction is usually less than 10\%.  Correcting
 for extinction and flux dilution
we derive distances to 10 supernovae: SN 1968L, SN 1969L, SN 1970G,
SN 1973R, SN 1979C, SN 1980K, SN 1987A, SN 1988A, SN 1990E, and SN 1990ae.
The distance measurements span
a wide range, 50 kpc to 120 Mpc, which is unique among the methods for
establishing the extragalactic
distance scale.  The distances measured to SN 1970G in M101 and SN 1987A in
the LMC are in good agreement
with distances determined from Cepheid variable stars.  Our distance to the
Virgo Cluster, $22\pm3$ Mpc,
is larger than recent distances estimates made using surface brightness
fluctuations, planetary
nebula luminosity functions, and the Tully-Fisher method.  Using the
distances determined from
these type II supernovae we derive a value of
$H_0 =  60 \pm 10$ km sec$^{-1}$Mpc$^{-1}$. This value is
subject to errors caused by local deviations in the Hubble flow, but will
soon be improved by applying the
Expanding Photosphere Method to several distant type II supernovae.

\noindent {\it Subject headings:} Cosmology: Observations - distance scale
- supernovae: general

\vskip .2 in
\centerline {\bf I. Introduction}
\vskip .2 in

The path to the extragalactic distance scale is long, complex, and
pitted with traps for the unwary.
Most distance indicators that are useful at the distance of the Virgo Cluster
and beyond are
calibrated with nearby galaxies whose distances have been measured using
Cepheids. This
bootstrapping process is far from ideal because the nearby galaxies used
for calibration may not represent
the distant galaxies where a method is applied. The ideal distance indicator
 would
be independent of distances in the Milky Way or to nearby galaxies, and
would employ the same method for both local and distant galaxies. Based
on a suggestion by Leonard
Searle, Kirshner and Kwan (1974) described a
method with these desirable properties, the Expanding Photosphere
Method (a.k.a.
Baade-Wesselink Method), which uses Type II supernovae (SN II)
as ``custom yardsticks". The Expanding Photosphere Method (EPM hereinafter)
is independent of every
part of the extragalactic distance ladder, and can be used to measure
distances to individual SN II
at 50 kpc, as well as 200 Mpc.  These attributes make SN II attractive
objects for
establishing the extragalactic distance scale.  In this paper we show
that EPM is accurate, widely
applicable, and has the potential to make a significant contribution to
the quest for $H_0$.

   The primitive form of EPM described by Kirshner and Kwan (1974)
assumes that a SN II
consists of a spherically symmetric expanding photosphere that radiates as
a blackbody. If $z<<1$, $$\theta= {R \over D }=\sqrt{{{f_\nu}\over{\zeta^2 \pi
B_\nu(T)}}},\eqno(1)$$
where $\theta$ is the angular size, $R$ is the radius of the supernova's
photosphere, $D$ is the distance to the supernova, $B_{\nu}(T)$ is the Planck
function, $f_\nu$ is the observed flux density of the supernova, and $\zeta^2$
is a
correction factor that accounts for the effects of flux dilution; for a
blackbody
$\zeta^2$=1.  The stellar envelope undergoes free expansion, and the
radius of the photosphere,
$R$, at time $t$, is just $$R = v(t-t_o)+R_o.\eqno(2)$$
The expansion velocity of the material that is instantaneously at the
photosphere, $v$, is determined from
absorption minima in the supernova spectra.  The radius of the supernova
progenitor at $t_0$, $R_o$,
is usually negligible since typically $ R > 10^{15}$cm and $R_0 << 10^{14}$cm.
Supernovae expand freely because their explosion energy is much greater than
the gravitational
binding energy of the progenitor stars. (Sakurai 1960; Grassberg, Imshennik,
and Nadyozhin 1971).
Deceleration by the interstellar medium and circumstellar gas is small
because the mass of matter
swept up by the supernova in its first months is much less than the mass
ejected
for any reasonable density of the circumstellar matter. The photosphere
always moves inward in mass
coordinates, and its radius is determined by the velocity of the material
which is
instantaneously at the photosphere. While the photosphere is receding in mass
coordinates, and may even be receding in spatial coordinates, it is still
being formed in material
that is freely expanding outwards.  Combining equations (1) and (2) we get
$$ t=D\left({{\theta}\over {v}}\right) + t_o.\eqno(3)$$
To determine $\theta$, the temperature of the photosphere must be measured
either from spectrophotometry or broad
band filter photometry.  This leaves two unknowns,
$D$ and $t_o$.  With two or more observations, separated by more than a
week, we can
determine the distance and time of explosion through a linear
regression.  If observations do not have sufficient separation, $\theta/v$
in equation (3) has a small range,
and it is difficult to determine accurately $D$ and $t_0$ from the regression
(Kirshner, Arp and Dunlap 1976).
If $t_o$ is known, each observation provides an independent measurement of
the distance. This provides a useful internal test:  if the method is
effective, the data will give
the same distance to the supernova at all times.
EPM has been implemented in this way on several SN II (SN 1969L and SN 1970G,
Kirshner and Kwan 1974; SN 1979C, Branch {\it et al.} 1983; SN 1979C and
SN 1980K, Kirshner 1985;
SN 1987A, Branch 1987, Jeffery and Branch 1990). In addition, Schurmann,
Arnett and Falk (1979)
modeled SN II light curves, and by comparing their shape and color to
observations, estimated the distance to
SN 1959D, SN 1968L, SN 1969L and SN 1970G.

This simple picture evades some significant challenges to obtaining
accurate distances from SN II.
SN II do not radiate as perfect blackbodies (Wagoner 1981, Shaviv, Wehrse,
and Wagoner 1985), and the blackbody correction factors, $\zeta^2$, must
be determined.
The effect of interstellar reddening and extinction, and the possibility of
asymmetric expansion
of supernova photospheres also need to be considered.  Several groups have
modeled SN 1987A's atmosphere
to determine blackbody correction factors (Chilukuri and Wagoner 1988,
H\"oflich 1988, Eastman and
Kirshner 1989, Schmutz {\it et al.} 1990). The distances determined to
SN 1987A using these
models, $44$ kpc $\simless D_{LMC}\simless$ 50 kpc, agree with the distance
given from Cepheids, $49.4 \pm 3.5$ kpc
(Walker 1987), and give encouragement that SN II
can provide reliable distances to galaxies.   In addition, Hauschildt,
Shaviv, and Wehrse (1989)
have used models of SN II atmospheres to estimate the distances to SN
1980K, and SN 1979C.

Here we develop methods for dealing with departures from blackbody emission
and with extinction from
interstellar dust, then use these results to measure distances to
10 supernovae.  Section II discusses atmospheric models and blackbody
correction factors.
Section III evaluates the effects of extinction in a distance measurement.
Asymmetric supernovae
and their role in distance determinations are briefly discussed in Section IV.
In Section V we
note the advantages of using infrared photometry to measure distances to
SN II. Section
VI contains a short summary of the data for each of the 10 supernovae, and
our distance measurements.
The distances derived using EPM are compared
to distances determined using other methods in section VII. In
section VIII we discuss our estimate of $H_0$.

\vskip .2 in
\centerline {\bf II. Flux Dilution and Models}
\vskip .2 in

SN II have scattering-dominated atmospheres: the flux emerging from their
photospheres is smaller than that of a blackbody with the same radius and
color temperature (Wagoner 1981,
Shaviv {\it et al.} 1985). We characterize this dilution of flux by
the quantity $\zeta^2$. The actual distance from equation (1), $D_{SN}$, is
then a factor of $\zeta$ less
than the apparent distance measured assuming a SN is a perfect blackbody
(i.e. $D_{SN}= \zeta D_{app})$.
Random walk arguments show that the radiation field of
an atmosphere is thermalized at an optical depth of
$$\tau_{therm}\approx \sqrt{{{\sigma+\kappa} \over {3\kappa}}},\eqno(4)$$
where $\sigma$ is the scattering opacity, and $\kappa$ is the absorptive
opacity.  An approximate solution to the radiative diffusion equation for a
static plane-parallel
atmosphere shows that the emergent flux is

 $$F_\nu={{4}\over{\sqrt{3}}}\left({1\over{1+{\sqrt{3}\tau_{therm}}}}\right)
\pi B_\nu(T_{therm}) = \zeta^2 \pi B_\nu(T_{therm}) , \eqno (5)$$
where $B_\nu(T_{therm})$ is the Planck function at the temperature where the
radiation
field is thermalized (Mihalas 1978; Rybicki and Lightman 1979; Hershkowitz,
Linder and Wagoner 1986a).
If $\sigma >> \kappa$, the emergent flux
will be much less than $\pi B_\nu (T_{therm})$, and
the emission will be that of a dilute blackbody. Figure 1 shows the opacity
computed for a 15$M_\odot$
red supergiant (RSG hereinafter) supernova by Woosley and Weaver (personal
communication) 10
days after core collapse; $\sigma$ is 70 times $\kappa$, and the flux should
be dilute by a
factor of 4.  Because supernova atmospheres are not plane-parallel, and the
ratio of $\sigma$ to
$\kappa$ varies with optical depth and wavelength, detailed models of
supernova atmospheres are essential
to get a reliable estimate of $\zeta$.

Motivated by SN 1987A, several groups have recently developed new codes (or
adapted existing ones)
to model SN II atmospheres which are typically time independent, spherically
symmetric, and employ non-LTE equations of statistical equilibrium for
hydrogen and helium
(Chilukuri and Wagoner 1988, H\"oflich 1988, Eastman and Kirshner
1989, Schmutz {\it et al.} 1990).  These models are successful at
reproducing the optical
spectrum of SN 1987A at several different epochs.  In addition, these
atmospheric models predict the total emergent flux
(and $\zeta$), and can be used to determine the distance to the LMC.  The
distances
determined to SN 1987A by these independent groups are consistent to about
$\pm10\%$, and agree
remarkably well with Cepheid and RR~Lyrae distances for the LMC (Table 1).
SN 1987A
resulted from the explosion of a B3 star, and models were crafted
specifically to fit that event.
Taking the next step and determining distances to more distant galaxies
requires models for typical
SN II, which likely result from exploding massive stars in the RSG phase.

The code we use to construct model atmospheres is an improved version of
that used by Eastman and
Kirshner (1989). These models are time-independent, but include the
effects of spherical geometry,
and treat all relativistic effects which are important in an expanding
gas to an accuracy of $v/c$.
The models can employ non-LTE
equations of statistical equilibrium for a large number of
atoms, but here only H and He are treated in
this manner. The excitation and ionization of all other elements is
computed from the Saha-Boltzman
equation.  These populations are used to approximate the opacity from
63,000 transitions in heavy
elements, most of which lie in the UV, and which have a major
effect on the UV radiation transport (Karp {\it et al.}
1977).

For each model the lower and upper boundary conditions must be specified,
as well as the velocity and density structure. The upper boundary condition
is given by the assumption that there is no radiation striking the
atmosphere from above.  The lower boundary condition is usually given as the
total luminosity at the lower boundary of the atmosphere.  The first
assumption could be in
error if there is a significant amount of circumstellar material. This
material, however, produces
a noticeable effect on the SN spectrum (see $\S$VI, SN 1979C) and should
not generally be an
undetected problem. The velocity
and density structure are generally supplied from hydrodynamic calculations
(e.g. Woosley 1988), but for convenience are sometimes parameterized
in the form of a power law where the  density structure is
$$\rho(v,t)=\rho_0\left({v\over{v_0}}\right)^{-\gamma}\left({t\over{t_0}}
\right)^{-3}.\eqno(6)$$
This parameterization of the supernova atmosphere agrees with the hydrodynamic
calculations, and allows for easy experimentation with density structures.

  Ideally, to determine the distance to a supernova, the distance
correction factor,
$\zeta$, is calculated from models. First, we fit the synthetic
spectrum's color temperature.  The temperature is derived by comparing
the broad band
colors of the calculated spectrum ($B-V$, $V-R$ etc.) determined using
analytic filter
functions (described in $\S$VI) to those of a blackbody.
In (5) $\tau_{therm}$ and $T_{therm}$ may be dependent on wavelength, and
this will cause supernovae spectra to deviate
from a Planck function.  Therefore determining the temperature of a
supernova over a broad range
of wavelength  (e.g. $(V-H)$) should be avoided.
We determine $\zeta$ from the ratio of the modeled luminosity of
the supernova in a given wavelength range to the luminosity
of a blackbody with radius equal to that of the photosphere at the fit
temperature.

Eastman and Kirshner (1989) modeled SN 1987A at 5 different epochs (1.85, 2.5,
5.5, 7.7, and 10.0 days).  Figure 2 shows $\zeta$, calculated from models
for both the optical $(4000$\AA$-6000$\AA)
and infrared $(1.2\mu - 2.2\mu )$, as a function of color temperature.
The optical distance
correction factors are large at early times when the supernova is
hot ($T_{opt} \approx 10000$~K), but become less important as the
supernova cools to
$T_{opt} \simless 6000$~K.  The corrections are much less important in the
infrared, where the
color temperature of the IR continuum is smaller than that of the optical.

We have calculated $\zeta$ for models of
15 $M_\odot$ and 25 $M_\odot$ RSG supernovae whose structures were
calculated by Woosley and Weaver
(personal communication).  These models
show that the evolution of $\zeta$ as a function of temperature is
similar to SN 1987A.  It would be a
great simplification if SN II atmospheres are sufficiently
similar that $\zeta$
depends only on the temperature of the photosphere. When the observations
of a SN II constrain
the value of $t_0$, it is possible to make empirical estimates of $\zeta$.
In this situation each observation provides an independent measure
of the distance to the SN, and
if $\zeta$ changes, so will the computed distance (Branch 1987).
Equation~(1) gives
$$\zeta_{emp}(T_{opt}) = {{D_{SN}}\over{D_{app}(T_{opt})}}, \eqno(7)$$
where $D_{SN}$ is the actual distance to the SN, and $D_{app}(T_{opt})$
is the distance determined to the SN at the optical color temperature
assuming $\zeta=1$.  If we take the distance to the LMC to be 50~kpc,
and the explosion time from the observation
of neutrinos (Bionta {\it et al.} 1987, Hirata {\it et al.} 1987),
we can derive $\zeta$
from the application of EPM on SN~1987A (see $\S$VI for discussion
of observations).
Figure~3 shows $\zeta(T_{opt})$ as measured from SN 1987A, and suggests that
$\zeta \approx 1$ when $T_{opt} \approx 5500$~K.  Our models also show
that $\zeta$ approaches 1 when
$T_{opt} \approx 6000$~K for SN II, however, we have not yet made
models with $T_{opt} < 6000$~K.
As the gas temperature at the photosphere decreases, the ratio of absorptive
to scattering opacity at optical wavelengths increases. The optical
scattering opacity is mainly electron
scattering, and is independent of temperature above the recombination
temperature of hydrogen.
The most important absorptive opacity source is photoionization out of
hydrogen $n=3$, which is
a sensitive function of $T$. As the temperature decreases the
fraction of H atoms
in $n=3$ goes up, and consequently, $\zeta$ increases.
A rough first order analysis suggests the distance correction factor
should vary with gas density as $\rho^{1/4}$.
The already low sensitivity
of $\zeta$ to $\rho$ might be softened somewhat further by other factors,
such as the effect which varying $\rho$ has on the temperature structure and
departure from thermal equilibrium. The details of this relationship will be
explored more thoroughly in a future paper.
At any given color temperature, the models and, presumably,
the supernovae shown in figure~4, span a range of
values for the gas density at the photosphere, which probably accounts for
much of the scatter in the figure. For the model calculations shown in
figure~4, the density at the photosphere at all times varied by less than
a factor of 10. Apparently, over the range of densities present at
the photospheres of SN II, $\zeta$ is most sensitive to temperature.

As the photospheric temperature approaches the hydrogen recombination
temperature, $T_{rec}$, a recombination wave begins moving inward through the
envelope. In this phase, the photosphere is always at the place where
hydrogen is recombining. The thermalization depth reaches a minimum near
$T_{rec}$, and the distance correction factor approaches unity.

If we assume that $\zeta=1$ when $T_{opt} < 5500$~K, $\zeta$ can be
estimated  for other SN II whose times of explosion are roughly
known, but whose distances are not.  Figure~4 shows $\zeta(T_{opt})$
as computed from our models, and measured empirically from
SN 1987A and four other SN II.   The agreement is remarkable,
especially considering the errors are at least $\pm$(10-20)\% for the
empirically derived
factors, and around $\pm10\%$ for the factors derived from the models.
The agreement also
supports the assumption that SN II are good blackbodies at 5500~K
(if they were not, the correction factors derived empirically and using
models would be offset
from each other).    The evidence that $\zeta$ depends uniformly on
temperature is persuasive,
but incomplete, and in the future we will model the
atmospheres of many more SN II to further test
this hypothesis. In the meantime we apply the results of Figure~4 to several
historical supernovae (about as old as the youngest author) and a few more
recent ones.

\vskip .2 in

\centerline {\bf III. Extinction}
\vskip .2 in
As the light from a supernova travels to the Earth, it passes though dust
both in the parent galaxy and in the Milky Way.  Dust reddens the light
and diminishes its intensity, and we expect these effects to
cause some uncertainty in applying EPM.  Extinction in the Milky Way has been
well studied (Burstein and Heiles 1982, 1984). However, there are still sizable
uncertainties in the extinction at low galactic latitude.  Extinction from
dust within
a supernova's host galaxy is much harder to quantify. SN II always occur
in galaxies rich with
dust, and the total extinction to a supernova may have large uncertainties,
especially if the galaxy
is a highly inclined spiral.  Problems with extinction are minimized
when a supernova
occurs at high galactic latitude and in the outer regions of a face-on spiral.
In these instances
the total extinction is expected to be small, and should not pose a
significant problem.

A direct method of determining the extinction
to a supernova is to measure the equivalent width of Na and Ca
interstellar absorption
against the continuum source provided by the supernova.  Although the
observations are
difficult and the interpretation is uncertain because of unknown dust to
gas ratio, the
measurement gives a direct measure of the column density of gas, and we
expect this to
be proportional to the extinction. This method has been applied to
several supernovae; e.g. Penston and Blades (1980) for SN 1979C,
Pettini {\it et al.} (1982) for SN 1980K, Blades {\it et al.} (1988)
for SN 1987A, and Steidel, Rich and McCarthy (1990) for SN 1989M.

SN II light curves are subdivided into two classes based on their shape;
linear (II-L),
and plateau (II-P). SN II-L, after maximum,
decline exponentially in brightness (linear in magnitudes),
whereas SN II-P, 25 days after
maximum, maintain a nearly constant brightness for several weeks
(Barbon, Ciatti and Rosino 1979;
Kirshner 1990). A less direct method of determining the extinction for
a SN II-P is to
examine its color evolution. SN II-P
start out hotter than 10000~K ($B-V \simless 0.2$), and cool to the
recombination temperature of hydrogen (5500~K) during the plateau phase.
During the plateau phase,
which lasts for several weeks, the supernova has a constant color of
$(B-V) = 0.75 \pm 0.1$ as shown below.
This evolution tightly constrains the amount of reddening present in a SN II-P.
The recombination temperature of hydrogen is relatively insensitive to
density and metallicity,
and should be nearly the same for all SN II-P.  This insensitivity of
the recombination temperature to density and
metallicity can be demonstrated by using Saha's equation to determine
when hydrogen ionizes, over the range of
densities indicated by hydrodynamic calculations of SN 1987A
(Woosley 1988) and 15$M_\odot$ and  25$M_\odot$ RSG supernovae. The
calculation, simple as it may be, yields a recombination
temperature ($5500~K \pm 500~K$) which is consistent with SN 1987A
(table 8), when
corrected for $A_V=0.6$ as determined by Blanco {\it et al.} (1987).
 It is also consistent with
SN 1968L (table 2b), SN 1969L (table 3b), SN 1990ae (table 11b).  All of
these supernovae occurred at high
galactic latitude;  in addition SN 1968L occurred in a face on galaxy,
and  SN 1969L and SN 1990ae occurred in the outer regions of
their parent galaxy, where we expect the total reddening is small.
Figure~5 shows the
($B-V$) evolution for SN 1968L, SN 1969L and SN 1990ae.  Two supernovae in
highly inclined spirals, SN 1973R and SN 1990E, are also plotted.  All of
the supernovae
have evolved in a similar manner, except that SN 1973R and SN 1990E
are shifted by
a constant ($B-V$). This shift represents the color excess, $E(B-V)$,
and indicates the
amount of extinction. Using $A_V = 3E(B-V)$ (Whitford 1958) to determine
the visual extinction suggests
that if the $(B-V)$ color excess can be determined to  $\pm 0.1$ magnitudes,
the extinction to SN II-P
can be estimated to $\pm 0.3$ magnitudes.

Surprisingly, even a moderately large extinction does not produce a
large effect
in the estimated distance.  EPM is affected in two ways by extinction.
A supernova appears dimmer, and hence we overestimate its distance.
However, a supernova
also appears redder and cooler, and
therefore we underestimate the distance.  In addition, because the
supernova appears cooler, the
distance correction factors ($\zeta$) from Figure 4 are closer to unity,
and we overestimate the
distance.  These effects can cancel each other to a remarkable extent. The
effect of extinction on the
distance determination has been calculated for several individual
supernovae (Figures 6 and 7).
The uncertainty in distance due to extinction is $< 10\%$ in most cases,
even though the
uncertainity in the extinction is $\pm 0.3$ magnitudes.  The
uncertainties, however,
can be much larger in cases where the supernova is not
observed continuously over a long period of time.  Extinction has a
much smaller effect in the
infrared.  If infrared photometry $(JHK)$ is used to measure a distance,
the distance is systematically underestimated by less than 10\% for a
full magnitude of
unaccounted visual extinction.

\vskip .2 in
\centerline {\bf IV. Asymmetries}
\vskip .2 in
Polarization (Cropper {\it et al.} 1988),  speckle observations
(Papaliolios {\it et al.} 1989)
and HST observations of the circumstellar shell (Jakobsen {\it et al.}
1991) give indications
for some degree of asymmetry in SN 1987A.  Although an asymmetric
photosphere could
affect the individual distance determinations, there are several
reasons to believe it is not a
serious problem for EPM.
The polarization measurements may indicate moderate asymmetries (20\%)
at early times in SN 1987A.  However, these determinations are hampered by
uncertainties in the polarization due to the interstellar medium which are as
large as the observed polarization in SN 1987A (Jeffery 1989).  If the
inferred asymmetry for
SN 1987A were correct, it would produce a 10\% uncertainty in the
derived distance to the
supernova (Wagoner 1991). The speckle observations indicate large
asymmetries in SN 1987A, but only several months after the explosion.
These observations
are looking at material substantially deeper inside the supernova than
the gas that formed
the photosphere during the first 50 days of SN 1987A.

Chevalier and Soker (1989) have shown that a supernova
arising from the explosion of a RSG, which we identify with a typical SN II-P,
is likely to have asymmetries much smaller than found in SN 1987A.  The
flatter density profile
of the RSG reduces the asymmetry of the propagating shock relative to
SN 1987A.  Additionally, because a RSG is more extended, rotationally induced
asymmetries will be smaller than for compact progenitors.  Most bright SN II
models require extended
RSG progenitors (Chevalier 1976) and should have smaller uncertainties from
asymmetries
than SN 1987A.

Wagoner (1991) estimated the effect of asymmetries on the
determined distance of a SN II.  He showed that although the individual
effects can be large for highly
flattened systems, the average distance determined, when summed over all
viewing
angles, is within 1\% of the correct distance.  Large asymmetries
could cause substantial errors in individual distance determinations, but
they would not
bias the derived global distance scale to smaller or larger distances.

\vskip .2 in
\centerline {\bf V. Using the Infrared}
\vskip .2 in
With the advent of infrared arrays, it is now possible to obtain infrared
photometry ($JHK$) of relatively faint supernovae.  The infrared
has several advantages over optical observations for applying EPM.
First, as shown in $\S$III, the uncertainty in a distance measurement
due to extinction is less than half that incurred when optical
photometry is used.
Second, the early spectra of SN 1987A (Elias
{\it et al.} 1988) show few spectral features in the J, and especially the
H and K bands, making it easier to determine the temperature from the
broad band colors.
Finally, as discussed previously, the flux dilution in the infrared is
much smaller.

Flux dilution depends on the ratio of the scattering opacity to the absorptive
opacity (equation (4) and (5)).  In the infrared, free-free absorption
begins to dominate the bound-free process, and is the major source of the
absorptive opacity. The opacity due to free-free transitions is proportional
to $\lambda^3$.  Thus, we expect SN II to better approximate blackbodies
at longer
wavelengths. This prediction is borne out in our models, which show the
blackbody correction factors near unity for near infrared wavelengths
(Figure 2).  This result is not consistent with the models of Hershkowitz,
Linder, and Wagoner (1986b) and Hershkowitz and Wagoner (1987), whose
plane-parallel models, which
treat hydrogen (6 levels) in non-LTE, show that at $T_{IR}>6000~K$, the
emergent flux
in the infrared is dilute for any reasonable density of the supernova
photosphere.  The effective temperature of a supernova photosphere is typically
larger than the color temperature, and this may account for the discrepancy.

	Measuring the color temperature of a supernova in the infrared is more
difficult than in the optical because the spectrum is close to Rayleigh-Jeans.
Infrared photometry is generally of lower accuracy than optical, and this
accentuates the problem.  Fortunately, in the Rayleigh-Jeans regime, the flux
and temperature are linearly related, and therefore the distance depends on
$\sqrt{T}$.  If the error in the $J$, $H$, and $K$ photometry is
$\pm.05$ magnitudes,
the uncertainty in the distance for a single measurement is 7\% at 5500~K, and
is 12\% at 10000~K.  SN~II-P cool to $T_{IR} < 5500$~K very quickly,
and remain at that temperature
while on the plateau phase. Therefore, determining the color temperature
from $(JHK)$ will
not cause significant errors except at the earliest stages in a supernova's
evolution.

\vskip .2 in
\centerline {\bf VI. Observations and Distance Determinations}
\vskip .2 in

    We present a summary of the observational data and its implementation
in making
distance determinations to 10 SN II.  The temperatures of the
supernovae are determined in one of two ways.  In most cases the color
temperature is
fit from the photometry using analytic filter functions.  For
$B$ and $V$ the functions of
A\u zusienis and Strai\u zys (1969) are used. The filter functions of
Bessel (1983) are used
for $R_C$ and $I_C$, but are shifted 23{\AA}  to the blue and 50{\AA}
to the red, respectively, as
suggested by Taylor (1986). The filter functions
of Johnson (1965) are convolved with the transmission through a
model terrestrial atmosphere
(Selby and McClatchey 1972) for the $J$, $H$, $K$, and $L$ bands.
The analytic filter
functions have been checked
by determining the broad band colors of the spectrophotometric standards
of Oke and Gunn (1983),
and they agree in $B$ and $V$ to $\pm 0.03$ magnitudes. Integrating
blackbodies with these
filter functions yields the relation between color temperature and
$(B-V)$ color:
$${{10^4 K}\over{T}} = 1.605(B-V)+.67. \eqno(9)$$
In other cases where photometry is not available, but flux-calibrated
spectra are, the temperatures
are fit directly from the spectra by comparing the continuum to a
blackbody.  The two methods generally
agree to 10\% in temperature (Figure 8), however temperatures
measured from spectra are
systematically $\approx$ 5\% smaller than those measured
from the $(B-V)$ color.
Systematic effects of this size have a minor effect on a distance
determination
(5\% to 10\% depending on the temperature).
In cases where both spectra and photometry are available, we have used
photometry.
In addition, all distances derived from optical photometry are corrected
for flux
dilution using the solid line in Figure 4.  The velocity
of a supernova's photosphere is usually derived from the absorption
minimum of the FeII lines
$\lambda5169$, $\lambda5018$, and $\lambda4924$ as suggested by the
calculations of Eastman and Kirshner (1989) and Schmutz {\it et al.}
(1990).  These lines
are narrow, and are formed at small optical depth for the first several
weeks in a
supernova's evolution. In a few instances, at early times, these
lines are not easily visible,
and the velocities are derived
from the Balmer lines.  Models, if they were available, could be used to
estimate the photospheric
velocity from the Balmer lines.  In practice, however, the
velocities measured from the hydrogen
lines are corrected by the ratio of the velocities of the
hydrogen and iron lines when the iron lines
are first visible. If there are periods where photometry is
available, and spectra are not,
velocities are estimated through linear interpolation or
extrapolation.  In all cases the internal scatter
of the distance measurements is less than 10\%. Our estimates of the
total error, which range from 5\% to 30\%
also include uncertainties in blackbody correction factors and reddening.

\vskip .2 in
\noindent {\bf SN 1968L}
\vskip .1 in
NGC 5236 has been the site of 6 supernovae this century. Theg fifth,
SN 1968L, was
discovered on 1968 July 16 near the nucleus by J.C. Bennett.
Photographic spectra and photometric observations were made by Wood and
Andrews (1974),
and further photoelectric photometry $(BV)$ was obtained by Wamsteker (1972).
Wamsteker's data were corrected for the brightness of the galaxy background by
Wood and Andrews (1974).  The data cover the first two months of the
supernova's
evolution quite thoroughly, and show it had a plateau light curve, but the
quality of both the spectra and photometry
is poor. The RMS errors in the velocities measured from spectra by
Wood and Andrews (1974)
are around 20\%, and the RMS error in the corrected BV photoelectric
photometry is
approximately 0.2 magnitudes (Table 2a).

To determine the temperature from the ($B-V$) light curve, we have corrected
for the foreground galactic extinction,  $A_V=0.11$
(Burstein and Heiles 1984). NGC 5236 is nearly face-on, and we see no evidence
for significant reddening of the supernova from the $(B-V$) color of the
plateau.
Because the individual spectra are poor, a mean velocity curve is used,
instead of
determining the velocities at each time.
Table 2b and Figure~9  show the results of the distance determination.  We find
a distance for NGC 5236 of $4.8^{+1.3}_{-0.7} $ Mpc.

\vskip .2 in

\noindent {\bf SN 1969L}
\vskip .1 in
SN 1969L was discovered on 1969 December 2 in outer regions of NGC
1058 as a $m{_B}$=13 object. Photographic spectra and photometry (BV)
were published by Ciatti, Rosino and Bertola (1971).  In addition Kirshner
and Kwan
(1974) published photographic and photoelectric spectra of the supernova.
 The data,
which are reasonably extensive and of good quality, show that SN
1969L was a type II-P, and was discovered near maximum (Table 3a).

The data are corrected for a foreground extinction of $A_V=0.17$ (Burstein and
Heiles 1984), and the temperature is calculated from the BV photometry.
SN 1969L should be free from significant extinction in the
parent galaxy, as it occurred 227$^{\prime \prime}$ from the nucleus of
the galaxy.  A distance of $11.2^{+1.0}_{-2.0}$ Mpc for SN 1969L is
derived from the
data (Table 3b and Figure~9).
\vskip .2 in
\noindent {\bf SN 1970G}
\vskip .1 in
SN 1970G was discovered in M101 (NGC 5457) on 1970 July 30. Photoelectric
photometry $(UBV)$ was published by Winzer (1974), and showed SN 1970G to be
a type II-L. Both photoelectric and
photographic spectra were published by Kirshner and Kwan (1974).
The early spectroscopic and photometric coverage of SN 1970G extends to only
thirty days after discovery, but it is of good quality (Table 4a).

There is little foreground extinction to M101 (Burstein and Heiles 1984).
However,
the supernova occurred in an HII region studied by Searle (1971), who
estimated the
total internal extinction to be $A_V=0.44$.  Since SN 1970G did not have
a plateau
phase, we cannot estimate the extinction from its color evolution, and we adopt
$A_V=0.44$ as the extinction to the supernova.  The temperatures
are derived from the BV photometry, except for days 35 and 70.
In these cases the temperature and flux are taken from spectrophotometry
published by Kirshner and Kwan (1974). The temperature measured from the
spectrophotometry is consistent (within the expected errors) with those
measurements made
from the photometry.  Our distance determination to M101 is
7.6$^{+1.0}_{-2.2}$ Mpc (Table 4b and Figure~9).

\vskip .2 in
\noindent {\bf SN 1973R}
\vskip .1 in
SN 1973R was discovered in NGC 3627 on 1973 December 19. Ciatti and Rosino
(1977) published photographic photometry (BV) and
spectroscopy of the supernova, and
Kirshner and Kwan (1975) published photoelectric spectrophotometry.
The photometric
coverage is quite good and showed SN 1973R to be a highly reddened type
II-P supernova. The spectral coverage of SN 1973R is limited to only
three epochs, but
data are of good quality (Table 5a).

NGC 3627 suffers from only a small amount of extinction from the Milky Way
(Burstein and Heiles 1984), however NGC 3627 is a very dusty spiral that is
inclined.
During the plateau phase, SN 1973R had $(B-V)=1.65$, which gives a color
excess of
$0.9\pm0.1$ magnitudes, and implies that $A_V=2.7\pm .3$ magnitudes.
The temperature and photometric angular size are
determined from the BV photometry.  The results are given in Table 5b
and in Figure~10, and give the distance to SN 1973R as
$7.6^{+2.2}_{-1.5}$ Mpc.

\vskip .2 in
\noindent {\bf SN 1979C}
\vskip .1 in
SN 1979C was discovered in M~100 (NGC 4321) on 1979 April 19.
Spectrophotometry is available
 from Branch {\it et al.} (1983) and
Kirshner (unpublished), and there is photographic spectroscopy
from Panagia {\it et al.}
(1980). Photoelectric photometry was published by de Vaucouleurs
{\it et al.} (1979), and
Barnes {\it et al.} (1979). Additional photographic photometry was
published by Ciatti
{\it et al.} (1979).  The spectral coverage of this supernova is
very good, and the photometric coverage is relatively good (Table 6).  SN 1979C
was an unusual SN II-L and was also observed extensively in the UV,
radio, and X-ray (Panagia {\it et al.} 1980). Fransson
(1982) has shown that the radio and UV observations suggest SN 1979C has
a significant
amount of circumstellar matter.

The extinction to SN 1979C was estimated by Penston and Blades (1980) and
Branch (1983) using
interstellar lines, and we adopt their value of $A_V=0.45$. The
temperatures are
derived from the BV photoelectric photometry, and a distance of $19 \pm 5$
Mpc is determined for
M100 (Table 7 and Figure~10). The large uncertainty is primarily due
to reddening (Figure 6).
This distance implies that SN 1979C was extremely luminous,
$M_B \approx -20.4$ (Table 12).

\vskip .2 in
\noindent {\bf SN 1980K}
\vskip .1 in
SN 1980K was discovered in NGC 6946 on 1980 October 29, and is the most
recent of the 6 observed
supernovae in this galaxy.  Extensive photographic $(UBV)$ and
photoelectric $(UBV)$ photometry were carried out by Barbon,
Ciatti and Rosino (1982) and Buta (1982) respectively.  In addition,
infrared photometry ($JHKL$) was
published by Dwek {\it et al.} (1983).  Extensive spectrophotometry
is also available from Uomoto and Kirshner (1986).  The data
for SN 1980K are exceptionally
good in coverage and quality, and show the supernova to be
type II-L (Table 7a).
SN 1980K was also observed in the UV (Panagia 1980), in the radio
(Sramek, van der Hulst and Weiler 1980),
and is still observable (Leibundgut {\it et al.} 1991).
In addition, a deep plate was taken of NGC 6946 49 days
before the discovery of the supernova.  From this plate
Thompson (1982) was able to
place an apparent magnitude limit of m$_F \simgreat 21.7$ for
the progenitor of SN 1980K.

NGC 6946 lies at low galactic latitude and has a foreground extinction of
$A_V=1.2$ (Burstein and Heiles 1984).  NGC 6946 is nearly face-on, and SN 1980K
occurred in the outer regions of the galaxy, so it is unlikely
that the supernova
was reddened substantially from the parent galaxy.  The
photometric angular size
was determined using both the infrared ($JHKL$) and the
optical (BV) photometry.
A distance of 8.1$\pm 1.5$ Mpc is derived from the infrared
observations alone, and the
optical data give an independent measure of $7.2^{+0.7}_{-1.0}$ Mpc
for the distance to NGC
6946 (Table 7b and Figure~10 ). This distance implies SN 1980K was
also quite luminous,
$M_B \approx -19.3$.
\vskip .2 in
\noindent {\bf SN 1987A}
\vskip .2 in
SN 1987A was discovered in the LMC on 1987 February 23.  It is one of the
most extensively observed extragalactic objects in the history of
astronomy - being observed
in wavelengths from the radio to gamma rays (Arnett
{\it et al.} 1989).  This paper
uses the photoelectric photometry $(UBVRIJHKLM)$ published
by Hamuy {\it et al.}
(1988) and Catchpole {\it et al.} (1987), and the spectral observations of
Phillips {\it et al.} (1987) (Table 8a).
In addition, the detection of neutrinos (Bionta {\it et al.} 1987,
Hirata {\it et al.} 1987) establishes
the time of explosion, and this is extremely useful as it eliminates one of the
two free parameters when applying EPM.  The knowledge
of the explosion date also acts as a test of the method if t$_0$
is left as a free parameter.

The total extinction to SN 1987A is determined to be $A_V=0.6$ (Blanco
{\it et al.} 1987).
Eastman and Kirshner (1989) have used specific models of SN 1987A during
its first 10 days to determine
a distance of $49 \pm 6$ kpc to the LMC. Individual models do not yet
exist for other SN II, so
here we will apply EPM to SN 1987A in a manner consistent with the
9 other supernovae.
We have used both optical $(VI_C)$ and infrared ($JHK$) photometry to determine
$\theta$ for the first 50 days of SN 1987A.  Line blanketing in the
$B$ band, which is easily observable
in spectra, makes the determination of the color temperature from
$(B-V)$ unrealistic for
SN 1987A, and therefore we use the temperatures derived from
$(V-I_C)$
when determining $\theta$ from the optical photometry.  For most
SN II $(V-I_C)$ gives color temperatures
which are systematically $\approx 500~K$ higher than $(B-V)$;
however, for SN 1987A the difference is $> 2000$K.
For the infrared, we calculate the color temperature
directly from the $JHK$ photometry.
With the IR data we assume that $\zeta = 1$ at all times
(as the models suggest), and
measure a distance to the LMC of 49$\pm$ 3 kpc (Table 8b and Figure~11 ).
If $t_0$
is left as a free parameter, the time of the
explosion using the IR data is derived
to be $0.7\pm0.7$ days after the neutrino event, so it appears that
EPM has performed well.
The optical data, corrected for flux dilution using figure~4,
yield a distance to the LMC of 53$\pm$4 kpc.  This distance is
shown for comparison; it is not an independent
determination of the distance because figure~4 is partially determined
using data for SN 1987A and
assuming a distance of 50 kpc to the LMC.

\vskip .2 in
{\bf SN 1988A}
\vskip .1 in
SN 1988A was discovered in M58 (NGC~4579) on 1988 January
15 near maximum light.
Visual, photographic, and photoelectric photometry (V) is published
by Ruiz-Lapuente {\it et al.} (1990) and Ruiz-Lapuente {\it et al.}
(1991). In addition there is photographic and CCD color
photometry from Benetti, Cappellaro, and Turatto (1991), and IAU
circulars (Kidger 1988, Binzel 1988, and Sadler and Simkin 1988).
Spectrophotometry is available from Ruiz-Lapuente {\it et al.} (1990),
Stathakis and Sadler (in preparation),
and Schlegel and Kirshner (in preparation) (Table 9a).  The photometric
coverage
is good but of marginal quality and shows SN 1988A to be type II-P.
The spectral coverage is limited,
but of good quality.

M58 has little foreground galactic extinction
(Burstein and Heiles 1984, Steidel, Rich, and McCarthy 1990), and
there is no evidence that there is a substantial amount of reddening from
the parent galaxy. The temperature evolution of SN 1988A is determined from
BV photometry, when available, and from spectra otherwise.  The data give
a distance to M58 of 23$\pm$4 Mpc (Table 9b and Figure~9).

\vskip .2 in
{\bf SN 1990E}
\vskip .1 in
SN 1990E was discovered near maximum light by the Berkeley Automated
Supernova Search on 1990 February 15.
In addition, it was not present down to $m_V=19$ on an image taken
on 1990 February 10 (Pennypacker and
Perlmutter 1990).  We have reduced
CCD Photometry (BVRI) taken with the 24$^{\prime\prime}$ telescope at FLWO
by R. Schild as well as
IR (JH) photometry on the same telescope by R. Peletier and S. Willner. In
addition we have used spectra reduced by B. Leibundgut and taken by J. Peters,
E. Horine, and A. Zabludoff on the 1.5m telescope at FLWO, as well as spectra
taken at the MMT. The details of the data will be discussed in
a future paper on SN 1990E. The spectral coverage of SN 1990E is very good,
but the data are of mediocre quality.  The optical photometry is of
good quality, but
the coverage is rather sparse.  The IR photometry coverage is sparse and of
poor quality ($0.25$ mag RMS).  The data show SN 1990E is a typical
type II-P supernova (Table 10a).

NGC 1035 lies at high galactic latitude and there is little foreground
extinction (Burstein and Heiles 1984), however NGC 1035 is a highly inclined
spiral, and therefore we expect SN 1990E to be reddened.  Comparing
the ($B-V$) color of SN 1990E with other
type II-P supernovae, the extinction to SN 1990E is estimated as
$A_V=1.0\pm.3$ (Figure 5). The optical temperature evolution for the
supernova is
determined from the BV photometry except for day 37, where it is
determined from a spectrum.
The IR temperature cannot be determined
with the JH photometry due to the poor quality of the data, and the
temperature measured from the optical spectrum is used. Our models
indicate that
this could overestimate the IR temperature by 15\%.  At these temperatures,
however, the flux in the infrared is directly proportional to the
temperature (Rayleigh-Jeans
regime) and the distance from this approximation
is overestimated by less than 11\%. The BV photometry gives a
distance for the supernova of $21^{+3}_{-3}$ Mpc and this is consistent
with the
distance derived from JH photometry of 22$\pm 6$ Mpc
(Table 10b and Figure~10 ).

\vskip .2 in
\noindent {\bf SN 1990ae}
\vskip .1 in

SN 1990ae was discovered well after maximum on 1990 October 15.  We have
two spectra
(Figure~12) of SN 1990ae.
The first was taken on 1990 October 19 by A. Zabludoff with the
MMT blue channel spectrograph
and reduced by B. Leibundgut. The second spectrum was obtained and
reduced by R.C. Smith
on 1990 November 12 with the MMT red channel spectrograph.  We also
have CCD photometry taken
with the KPNO 2.1m by A. Porter and B. Leibundgut on 1990 October
23, and 1990
November 9.  These data were reduced
within IRAF
using standard procedures. The quality of the data is superb (Table 11a).

There is little foreground extinction to SN 1990ae (Burstein and Heiles 1982),
and the supernova occurred in the outer regions of its parent
galaxy. Therefore we adopt $A_V=0.0$.  The temperature and photometric
angular size at both times were determined from the BV photometry.  The
distance
derived is $117^{+25}_{-15}$ Mpc as given in Table 11b.

\vskip .15 in
\centerline {\bf VII. Discussion of Distances}
\vskip .15 in

Several of the galaxies that contain SN II for which we have determined
distances have had their distances measured by other methods.  Most
notable of these are the LMC and M~101 for which Cepheid distances
are available, and which contained SN 1987A and SN 1970G respectively.

Cepheid and RR Lyrae stars have long been studied in the LMC
(e.g. Leavitt 1912),
and its distance is well established.  Our determination of the
distance to the LMC of 49$\pm3$ kpc using EPM
with infrared photometry, and 53$\pm 4$ kpc using $VI_C$ photometry, is in
excellent accord with the distance to
the LMC derived recently by Walker and Mack (1988) with RR Lyrae Stars,
48.8 kpc, and Walker (1987) with Cepheids, 49.4 kpc. It is also consistent
with other determinations of the distance to the LMC which use detailed
models of SN 1987A
to determine the flux dilution
(Chilukuri and Wagoner 1988, H\"oflich 1989,
Eastman and Kirshner 1989, Schmutz {\it et al.} 1990).
A novel method of determining the distance to SN 1987A which exploits
the geometry of the circumstellar ring observed with the Hubble
Space Telescope (Panagia {\it et al.} 1991)
gives a distance of $51\pm 3$ kpc which is also consistent with our results.

Cook, Aaronson and Illingworth (1986) found two Cepheids
in M101 at $m_R>23$.  These are the most distant Cepheids yet discovered and
give a distance of $7.1\pm.3$ Mpc for M101.  The distance to
M101 determined from
SN 1970G, $7.6^{+1.0}_{-2.2}$ Mpc, is in excellent accord with this Cepheid
distance. SN 1970G and SN 1987A are radically different from each other, with
different light curves, and different absolute magnitudes (Table 12),  yet EPM
has given a distance consistent with Cepheid
distances in both instances.  We believe this agreement with
Cepheid distances gives
strong support to the distances measured with EPM to other supernovae.

We can compare our results to several other methods of measuring
extragalactic distances.
The 21 cm line width-luminosity method of distance determination for spirals
pioneered by Tully and Fisher (1977)
has been widely used in the past 15 years (Aaronson, Huchra and Mould 1979,
Aaronson {\it et al.} 1982, Aaronson {\it et al.} 1986, Pierce and Tully 1988,
Fouqu\' e {\it et al.} 1990).  The method  has been calibrated using M31, M33,
and NGC 2403, and has been applied to hundreds
of spiral galaxies. The method works to relatively large distances
($cz \approx 7000$ km sec$^{-1}$), but gives
distances with typical errors greater than 15\%.
Recently, two additional methods of extragalactic distance determination
have been introduced which show great promise. The first uses
the planetary nebula luminosity function as a standard candle
(Jacoby and Ciardullo 1988, Jacoby
1989, Jacoby {\it et al.} 1989, Ciardullo {\it et al.} 1989a,
Ciardullo {\it et al.} 1989b,
Jacoby, Ciardullo, and Ford 1990). By cataloging planetary nebulae
in early type galaxies,
a luminosity function for planetary nebulae in a galaxy is constructed.
The luminosity function is calibrated with M31 and its distance using
Cepheids.
This method appears to have high internal precision for relative distances.
The method has been applied to the LMC and M81, and the derived
distances agree with
the Cepheid distances to both these galaxies.  It is worth noting
that M31, M81 and the LMC are
{\bf not} early type galaxies, and that this method provides
only distances relative to the
calibrator, M31. Recently, however, Ciardullo, Jacoby and
Harris (1991) have given evidence
that planetary nebulae
luminosity functions may be insensitive to the age of the
stellar population and galaxy Hubble type.
A second method uses the
surface brightness fluctuations of early type
galaxies as a distance indicator (Tonry and
Schneider 1988, Tonry and Schechter 1990, Tonry, Ajhar and Luppino 1990,
Tonry 1991).
The fluctuations are inversely proportional to distance, and relative distances
can be derived by taking the ratio of the surface brightness fluctuations of
two galaxies.
Absolute distances are measured by determining the fluctuations of a galaxy
whose distance is known by other methods - in this case the Cepheid distances
of M31 and M32.  Tonry (1991) has shown that there is a color effect due to
different
stellar populations in different galaxies. This calibration is not yet
complete, and could
be a source of error when using surface brightness fluctuations
to measure relative
distances.  The initial results, however, also show excellent
internal consistency.

NGC 5236, which is considered part of the Centaurus A group, has very
few measures of its own distance.  However, Centaurus A (NGC 5128) and the
group itself are well studied.
By fitting planetary nebula luminosity functions,
Jacoby {\it et al.} (1988) have determined the distance to the NGC
5128 as $3.8^{+.1}_{-.2}$ Mpc. Surface brightness fluctuations
(Tonry and Schechter 1990) give a somewhat smaller distance of $3.1\pm.1$ Mpc.
When comparing the distances to Centaurus A derived using the two
above methods with the distance
determined using SN 1968L in NGC 5236 of $4.8^{+1.3}_{-0.7}$ Mpc, one
must take into
account that the cluster depth may be a sizable fraction of the
distance to the cluster.  In this case it appears EPM is consistent
with the planetary nebulae distance, but is not consistent with the
surface brightness
fluctuation distance.

SN 1988A (NGC 4579) and SN 1979C (NGC 4321) both occurred in
galaxies that are associated
with the Virgo Cluster.   The Virgo Cluster is a good place to
compare the results of EPM
with other methods.  Jacoby {\it et al.}
(1990), using the planetary luminosity functions of 6 cluster members, have
determined a distance of 14.7$\pm1$ Mpc.  Tonry (1991),
using surface brightness fluctuations, derives a distance of 15.9$\pm0.9$
Mpc.  Applying the Tully-Fisher method to several galaxies of the Virgo
Cluster has typically given distances near 15 Mpc ( Aaronson {\it et al.} 1986,
Pierce and Tully 1988), however distances over 19 Mpc have been determined
by other groups (Fouqu\' e {\it et al.} 1990., Sandage and Tammann 1984).
Sandage and Tammann (1990), using the mean of 6 independent distance indicators
(globular clusters, novae, type Ia supernovae, $D_n$-$\sigma$,
Tully-Fisher, disk sizes
of Virgo spirals) derive a distance of 21.9$\pm.9$ Mpc.
Bartel (1991), using VLBI
observations to measure the angular radius of SN 1979C's radiosphere,
measured the distance
to M100 to be $22^{+7}_{-6}$ Mpc.  EPM, using SN 1988A ($23\pm 4$ Mpc)
and SN 1979C ( $19 \pm 5$ Mpc),
gives a distance of $22 \pm 3$ Mpc.  Since the depth of the Virgo Cluster
 is small compared
to the discrepancies, there appears to be a significant difference
between the two distance
scales.  If a distance of 15 Mpc were correct, it would require that
$\zeta < 0.7$ for SN 1988A
at all times we apply EPM. We believe $\zeta=0.9 \pm 0.1$ when
$T_{opt}\simless 6000$K. We have not yet matched
individual models to the spectra of SN 1988A, so we cannot
conclusively rule out the possibility that
$\zeta$ remains much less than unity even when the supernova cools
to below 6000~K, but it would
be a very surprising result.

SN 1990E occurred in NGC 1035, a highly inclined spiral for
which Aaronson {\it et al.} (1982)
have determined a distance relative to the Virgo Cluster using
the infrared Tully-Fisher relation.
Their distance, $D(NGC$ $1035)/D(Virgo)=.88$, is consistent with
our determination of $21^{+3}_{-3}$ Mpc
($D(NGC$ $1035)/D(Virgo)=.95$, comparing SN 1990E to  SN 1988A and SN 1979C).

The agreement between EPM and other methods is not yet satisfactory.
Although they generally
agree nearby (e.g. the LMC), they most certainly do not agree on the
distance to the Virgo
Cluster.  More cases need to be studied to sort out the discrepancies.
High quality
infrared data of future SN II-P in the Virgo Cluster will allow
EPM to give a firmer distance
estimate to the cluster.

\vskip .2 in
\centerline {\bf VIII. Determining $H_0$}
\vskip .2 in
The primary goal of determining the extragalactic distance scale
is to calculate $H_0$.  This is a difficult task even if local
distances are known exactly, due to local and large scale perturbations
in the Hubble flow
(e.g. Aaronson {\it et al.} 1982; Lynden-Bell {\it et al.} 1988).
EPM has the potential for circumventing
this problem because it works to large distances where perturbations
are expected to be small compared to the Hubble flow.
Unfortunately, distant SN II
($cz > 5000$ km sec$^{-1}$), which are relatively faint ($m_V > 18.0$), but
still easy to study with large telescopes, have not been considered
interesting objects
in the past, and little data has been
garnered on them.  In the future we plan to concentrate our
observational efforts on
SN II that occur beyond the Virgo Cluster.
The current sample permits us to estimate $H_0$ from
one distant supernova (SN 1990ae), and 8 supernovae which have occurred in
nearby galaxies.

      Determining $H_0$ from SN 1990ae is straightforward as its
velocity of recession, 7680 km sec$^{-1}$, is sufficient so that
velocity perturbations
should be less than a 10\% effect. Derived from this single galaxy,
ignoring all peculiar motions,
$H_0=66^{+9}_{-12}$ km sec$^{-1}$Mpc$^{-1}$.

      It is possible to use nearby galaxies to measure $H_0$ by comparing our
distances with velocities corrected for infall into Virgo and the
rotation of our galaxy (assumed to be
300 km sec$^{-1}$). We have created a Virgo infall
model using the equations set out by Schechter (1980) and
Kraan-Korteweg (1986).  This model has
four free parameters, $V_{Virgo}$, $V_{infall}$, $\gamma$, and $\delta
\rho/\rho$. Here
$V_{Virgo}$ is the observed velocity of recession for the Virgo Cluster
corrected for the
motion of our galaxy, $V_{infall}$ is the
rate that the Local Group is falling into the Virgo Cluster,
$\gamma$ is the power law index to the
virgocentric density profile (i.e. $\rho(r)=\rho_0r^{-\gamma}$),
and $\delta \rho/\rho$ is the
overdensity at the Local Group relative to the background density.
The values of these parameters
are not well agreed upon, and we choose ranges for these parameters
from a review by Huchra (1988) and
from Sandage and Tammann (1990). These values are:
900 km sec$^{-1} < V_{Virgo} < 1225$ km sec$^{-1}$,
200 km sec$^{-1}$$< V_{infall} <$375 km sec$^{-1}$, 2$< \gamma <$3,
and 2$< \delta \rho/\rho <$4.

The use of a virgocentric infall model significantly reduces the scatter
when trying to
determine $H_0$.  Fortunately, the exact choice of parameter values input
into the model
has a surprisingly small effect on the determination of $H_0$ (Table 12).
Figure~13
and Figure~14 show distance versus recession velocity, corrected for
Virgo infall, in the manner of
the original Hubble diagram (Hubble 1929). From Figure~13, which does
not include SN 1990ae, we derive
$H_0=58\pm 10$ km sec$^{-1}$ Mpc$^{-1}$.  From Figure~14, which includes
SN 1990ae,
we derive $H_0=60\pm 10$ km sec$^{-1}$ Mpc$^{-1}$.  This figure clearly
demonstrates the
linearity of the Hubble flow over a factor of 25 in distance.

      A less direct method of determining the Hubble constant is to
calibrate the absolute magnitude of type Ia supernovae.  There is
evidence that Type Ia supernovae are standard candles at maximum
(Hamuy {\it et al.} 1991, Miller and Branch 1990, Leibundgut 1991).
A Hubble diagram can be constructed from these standard candles which
is relatively free from local perturbations in the Hubble flow because SN
Ia have been observed in all
directions, and to large distances, $cz \approx 6000$ km sec$^{-1}$
 (e.g. Tammann and Leibundgut 1990).
If the absolute magnitude of type Ia supernovae were known, then $H_0$
could be determined.
Although this is not strictly an independent method of determining $H_0$,
it is a method to circumvent
uncertainties caused by local deviations of the Hubble flow.

Two of the galaxies for which we have measured the distance using EPM
have also
had a well observed SN Ia. SN 1989B, a SN Ia,  occurred in NGC 3627 16
years after SN 1973R.
Both supernovae were highly reddened, but the
reddening to both can be estimated.
Barbon {\it et al.} (1990) determined that SN 1989B
reached $m_B$=12.5 at maximum.  They also estimated the extinction
to be $A_B = 3.24$
by comparing the $(B-V)$ evolution of SN 1989B to a standard
color curve for SN~I by
Cappellaro (1982).  Wells {\it et al.} (in preparation) determine
$M_B^{max}=12.36$
for SN 1989B.  They derive $A_B=1.28$ by comparing the BVRIJHK colors of
SN 1989B to those of SN 1980N (Hamuy {\it et al.} 1990).  We favor the
determination of the reddening of Wells {\it et al.} because
the standard color curve by Capellaro is based on photographic photometry
of a few SN~I, and is much bluer than indicated by recent
photoelectric and CCD SN
photometry (e.g. 1980N and 1981D, Hamuy {\it et al.} 1990;
1990N, Leibundgut {\it et al.} 1991; 1991T, Phillips {\it et al.} 1992).
Our distance of 7.6 Mpc to SN 1973R, combined with the photometry and
reddening estimate from Wells {\it et al.},
implies an absolute maximum for SN 1989B of $M_B^{max}=-18.3$.  The
associated error due to reddening is too
large to allow calibration of SN Ia's from SN 1989B.  Uncertainty in
extinction causes equal uncertainties
in distances measurements when employing standard candles; however,
this is not the case with EPM.

SN 1989M, similarly, was found less than 2 years after SN 1988A in
NGC 4579.  SN 1989M
reached a maximum of $m_B$=12.2 (Kharadze and Pskovsky
1989).  Steidel, Rich, and McCarthy (1990) used SN 1989M to study the ISM
of NGC 4579 and measured very little extinction to the supernova.
Using our distance
of 23 Mpc derived from SN 1988A, an absolute magnitude at maximum for
SN 1989M is determined
to be $M_B^{max}=-19.6\pm 0.6$.  Combining a Hubble diagram, constructed
by Tammann and
Leibundgut (1990) using 35 type Ia supernovae at distances extending
to the Coma Cluster,
with our calibration of the absolute magnitude of type Ia supernovae
($M_B^{max}=-19.6\pm0.6$),
yields $H_0=51^{+19}_{-14}$ km sec$^{-1}$Mpc$^{-1}$.

Clearly these methods of determining $H_0$ are not perfected, even
if the distances derived using EPM were exactly correct.
However, there are many distant SN II discovered each year (8 in 1990
with $v_{rec} > 1500$ km sec$^{-1}$), for
which it should be possible to determine distances. In the next few
years, as we concentrate on gathering
data for faint SN II, a global measurement of $H_0$ free
from uncertainties due to local perturbations in the Hubble flow should be
feasible. The ability to measure accurate individual distances of
galaxies well beyond the Virgo
Cluster is an inherent advantage of EPM.  Planetary nebulae and
surface brightness fluctuations are limited to galaxies closer than 40 Mpc.
These two methods, however, do have outstanding internal consistency, and can
be applied at will to many nearby galaxies.  A more satisfactory
agreement with these methods
may result from further development of all three.
\vskip .2 in

\centerline {\bf IX. Conclusions}

We have explored the principal sources of systematic errors
that might occur when applying EPM.  We have presented distance correction
factors
based on radiative transfer in a scattering atmosphere from
models of SN 1987A and have shown that these corrections are
substantial. We have also found
that the correction factors are substantially less important at longer
wavelengths, and as
the supernova cools.  Modeling individual supernovae is very difficult,
but is the
ideal method of determining the distances to SN II.  However, preliminary
models for generic
25 $M_\odot$ and 15 $M_\odot$ RSG supernovae show the same evolution of
correction factors as SN 1987A, and suggest that blackbody correction
factors for most
SN II are dependent only on temperature, not on the details of the
individual progenitor.
Additional empirical evidence further supports this relation.  This
relation in its current
form (Figure 4) is very useful, but models need to be calculated
for a wide variety of
progenitors to improve our understanding of the correction factors.
The uniform
evolution of $(B-V)$ (i.e. their constant temperature on the plateau)
in SN II-P permits
a simple estimate of the total reddening to this type of supernova.
We have also explored the
effect of extinction on distance determinations and have shown that due
to canceling effects,
extinction has a surprisingly small effect (less than 10\% for 0.3 magnitudes
uncertainty in $A_V$) on the distance measurements.  Typical SN II should
not be prone to large
asymmetries at early times.  Even if asymmetries do exist in individual SN II,
the derived distance scale, on average, is not biased to small or
large values.  Much of the
uncertainty with EPM can be circumvented by working in the near
infrared. Our modeling indicates
that effects of extinction and flux dilution cause only small
uncertainties at wavelengths
longer than 1$\mu$.

We have applied blackbody correction factors derived empirically and from our
models to 10 supernovae, and have determined their distances. In addition, we
have used infrared photometry to determine independent distances to three of
these
supernovae.  Of these supernovae, SN 1987A and SN 1970G lie in
galaxies with Cepheid
distances.  In both cases the derived distances agree with the Cepheid
distances within the errors. In general, EPM gives slightly smaller
distances (15\%) than those proposed
by Sandage and Tammann (1990)  to other galaxies, and gives larger
distances ($\approx 50$\%) than
those derived using Tully-Fisher, planetary nebulae luminosity
functions and surface brightness fluctuations.

The distances derived in this paper, spanning the range from 50 kpc to
120 Mpc, are used to
estimate $H_0$.  Using the distant SN II, SN 1990ae, we
find $H_0=66^{+9}_{-12}$
km sec$^{-1}$Mpc$^{-1}$.
A Virgo infall model combined with the distances of the 10 supernovae
gives $H_0=60\pm10$ km sec$^{-1}$Mpc$^{-1}$.
Finally, we determine the absolute magnitude of type Ia
supernovae from two galaxies that have had both a well observed type Ia and
type II supernova.  These calculations yield $M_B^{max}=-19.6\pm 0.6$ for type
Ia supernovae.  This result, when combined with a Hubble diagram of distant
type Ia supernovae at maximum, yields
 $H_0=51^{+19}_{-14}$ km sec$^{-1}$Mpc$^{-1}$

Obtaining data for SN II is difficult because telescope time is scheduled
in advance, and supernovae are not.  Spectra need be taken only once a
week, because the
velocity of the SN photosphere does not evolve rapidly, except at early
times. $U$ and $R$ photometry are {\bf not}
as useful as $B$, $V$, $I$, $J$, $H$, $K$, and $L$ photometry because
the continuum is strongly contaminated by
absorption and emission in these two bands.  Special emphasis should be
placed on
obtaining infrared photometry because of the reduced uncertainty from
extinction and flux
dilution.  The observations should span from discovery until the
supernova leaves the plateau phase for
SN II-P, or about 50 to 100 days after maximum for SN II-L. After this
time the photosphere is no
longer formed within the hydrogen envelope, and loses all semblance to
a blackbody.
Observations while a supernova is young constrain $t_0$, and are
therefore especially valuable.

Applying EPM to supernovae is already a very useful method
of measuring extragalactic distances. In the future, by obtaining
high quality data of the many SN II discovered each year,
it will be possible to derive independent distances to many supernovae, most
of which will be well in the Hubble flow.  In this way we believe that
type II supernovae will become central to determining the
extragalactic distance scale.

We would like to thank Robert Wagoner for his careful reading of the
manuscript.  We would also like to thank Bruno Leibundgut, Phil Pinto,
and Mark Phillips for many
helpful conversations, and Stan Woosley and Tom Weaver for their models.
We are indebted to
Rudy Schild, Reynier Peletier, Steve Willner, Ann Zabludoff, Chris Smith,
Ed Horine, and Jim Peters for their willingness to take data on supernovae.
  This
research was supported in part by the United States National Science
 Foundation through
grant AST89-05529, and by NASA through grants NAGW-1273 and NAGW-2525.

\vfill
\eject

\centerline {\bf References}
\vskip .1 in

  \refitem { Aaronson, M. {\it et al.} 1982, {\it Ap.J.Supp.}, {\bf 50}, 241.}

  \refitem { Aaronson, M., Bothun, G., Cornell, M. E., Huchra, J. P.,
Schommer, P. A. 1986, {\it Ap.J.}, {\bf 302}, 536.}

  \refitem { Aaronson, M., Huchra, J., Mould, J. 1979, {\it Ap.J.},
{\bf 229}, 1.}

  \refitem { Arnett, W. D., Bahcall, J. N., Kirshner, R. P.,
Woosley, S. E. 1989, {\it Ann. Rev. Astron. Astrophys.}, {\bf 27}, 629.}

\refitem { A\u zusienis, A., Strai\u zys, V. 1969, {\it Sov. Ast.-A.J.},
{\bf 13}, 316.}

  \refitem { Barbon, R., Benetti, S., Cappellaro, E., Rosino, L.,
Turatto, M. 1990, {\it Astr.Ap.}, {\bf 237}, 79.}

  \refitem { Barbon, R., Ciatti, F., Rosino, L. 1979, {\it Astr.Ap.},
{\bf 72}, 287.}

  \refitem { Barbon, R., Ciatti, F., Rosino, L. 1982, {\it Astr.Ap.},
{\bf 116}, 35.}

  \refitem { Barnes, T., Beardsley, B., Moffett, T., Odewahn, W.,
de Vaucouleurs, G., de Vaucouleurs, A. 1979, {\it I.A.U. Circular 3386}.}

\refitem { Bartel, N. 1991 in {\it Supernovae, Proceedings of the Tenth
Santa Cruz Summer Workshop.}, ed. S. E. Woosley, (New York:
Springer-Verlag 1991) p. 760.}

\refitem { Benetti, S., Cappellaro, E., Turatto, M. 1991, {\it Astr.Ap.},
 {\bf 247}, 410.}

  \refitem { Bessel, M. S. 1983, {\it P.A.S.P.}, {\bf 95},480.}

  \refitem { Binzell, R. P. 1988, {\it I.A.U. Circular 4542}.}

  \refitem { Bionta {\it et al.}  1987, {\it Phys. Rev. Lett.}, {\bf 58},
 1494.}

  \refitem { Blades, J. C., Wheatley, J. M., Panagia, N., Grewing, M.,
Pettini, M., Wamsteker, W. 1988, {\it Ap.J.Lett.}, {\bf 332}, L75.}

  \refitem { Blanco V. M., {\it et al.} 1987, {\it Ap.J.}, {\bf 320}, 589.}

  \refitem { Branch, D. 1987, {\it Ap.J.Lett.}, {\bf 320}, L23.}

  \refitem { Branch, D., Falk, S. W., McCall, M. L., Rybski, P.,
 Uomoto, A. K., Wills, B. J.  1983., {\it Ap.J.}, {\bf 244}, 780.}

  \refitem { Burstein, D., Heiles, C. 1982, {\it A.J.}, {\bf 87}, 1165.}

  \refitem { Burstein, D., Heiles, C. 1984, {\it Ap.J.Supp.}, {\bf 54}, 33.}

  \refitem { Buta, R. 1982, {\it P.A.S.P.}, {\bf 94}, 578.}

  \refitem { Catchpole, R., {\it et al.} 1987, {\it M.N.R.A.S.},
 {\bf 229}, p.15.}

  \refitem { Chevalier, R. A. 1976, {\it Ap.J.}, {\bf 207}, 812.}

  \refitem { Chevalier, R. A., Soker, N. 1989, {\it Ap.J.}, {\bf 341}, 867.}

  \refitem { Chilkuri, M., Wagoner, R. V. 1988, in {\it Atmospheric
 Diagnostics of Stellar Evolution, IAU Colloquium 108}, ed. K. Nomoto,
 (Berlin: Spinger-Verlag) p. 295.}

  \refitem { Ciardullo, R., Jacoby, G. H., Ford, H. 1989a, {\it Ap.J.},
 {\bf 344}, 715.}

  \refitem { Ciardullo, R., Jacoby, G. H., Ford, H., Neill, J. D. 1989b,
{\it Ap.J.}, {\bf 339}, 53. }

 \refitem { Ciardullo, R., Jacoby, G. H., Harris, W. E. 1991, {\it Ap.J.},
 {\bf 383}, 487.}

  \refitem { Ciatti, F., Rafanelli, P., Ortolani, S., Rosino, L. 1979,
{\it I.A.U. Circulars 3361 and 3371}.}

  \refitem { Ciatti, F., Rosino, L. 1977, {\it Astr.Ap.}, {\bf 56}, 59.}

  \refitem { Ciatti, F., Rosino, L., Bertola, F. 1971,
{\it Mem. Soc. Astron. It.}, {\bf 42}, 163.}

  \refitem { Cook, K. H., Aaronson, M., Illingworth, G. 1986,
{\it Ap.J.Lett.}, {\bf 301}, L45.}

  \refitem { Cropper, M. Bailey, J., McCowage, J. Cannon, R. D., Couch,
W. J., Walsh, J. R., Strade, J. O., Freeman, F. 1988, {\it M.N.R.A.S.},
 {\bf 231}, 695.}

  \refitem { Dwek, E. {\it et al.} 1983, {\it Ap.J.}, {\bf 274}, 168.}

 \refitem { Eastman, R. G., Kirshner, R. P. 1989, {\it Ap.J.}, {\bf 347}, 771.}

  \refitem { Elias, J. H., Gregory, B., Phillips, M. M., Williams, R. E.,
 Graham, J. R., Meikle, W. P. S., Schwartz, R. D., Wilking, B. 1988,
{\it Ap.J.Lett.}, {\bf 331}, L9.}

  \refitem { Fouqu\' e, P., Bottinelli, L., Gouguenheim, L., Paturel, G.
1990, {\it Ap.J.}, {\bf 349}, 1.}

 \refitem { Fransson, C. 1982, {\it Astr.Ap.}, {\bf 111}, 140.}

  \refitem { Grassberg, E. K., Imshennik, V. S., Nadyozhin, D. K. 1971,
 {\it Ap. and Space Sci.}, {\bf 10}, 28.}

  \refitem { Hamuy, M., Phillips, M. M., Maza, J., Wischnjewsky, M.,
Uomoto, A., Landolt, A. U., Khatwani, R. 1991, {\it A.J.}, {\bf 102}, 208.}

  \refitem { Hamuy, M., Suntzeff, N. B., Gonzales, R., Martin, G. 1988
{\it A.J.},{\bf 95}, 63.}

  \refitem { Hauschildt, P. H., Shaviv, G., Wehrse, R. 1989, {\it Astr.Ap},
{\bf 210}, 262.}

  \refitem { Hershkowitz, S., Linder, E., Wagoner, R. V. 1986a, {\it Ap.J.},
 {\bf 301}, 220.}

  \refitem { Hershkowitz, S., Linder, E., Wagoner, R. V. 1986b, {\it Ap.J.},
 {\bf 303}, 800.}

 \refitem { Hershkowitz, S., Wagoner, R. V. 1987, {\it Ap.J.}, {\bf 322}, 967.}

  \refitem { Hirata, K., Kajita, T., Koshiba, M., Nakahota, M.,
Oyama, Y. 1987, {\it Phys. Rev. Lett.}, {\bf 58}, 1490.}

  \refitem { H\" oflich, P. 1988, in {\it Atmospheric Diagnostics of
Stellar Evolution, IAU Colloquium 108}, ed. K. Nomoto, (Berlin:
Spinger-Verlag) p288.}

  \refitem { Hubble, E. 1929, {\it Proc.Nat.Acad.Sci}, {\bf 15}, 168.}

  \refitem { Huchra, J.P. 1988, in {\it Proceedings of the A.S.P.: The
Extragalactic Distance Scale}, ed. S. van den Bergh and C.J. Pritchet,
(Provo: Brigham Young University Press) p 257.}

  \refitem { Jacoby, G., Ciardullo, R. 1988. in {\it Proceedings of the
1988 ASP Meeting on the Extragalactic Distance Scale}, ed. S. van den
Bergh and C. Pritchet, (San Francisco: Astronomical Society of the Pacific).}

  \refitem { Jacoby, G. H. 1989, {\it Ap.J.}, {\bf 339}, 39.}

  \refitem { Jacoby, G.H., Ciardullo, R., Ford, H., Booth, J. 1989,
{\it Ap.J.}, {\bf 344}, 704.}

  \refitem { Jacoby, G. H., Ciardullo, R., Ford, H. C. 1990, {\it Ap.J.},
{\bf 356}, 322.}

 \refitem { Jacoby, G. H., Walker, A. R., Ciardullo, R. 1990,
{\it Ap.J.}. {365}, 471.}

  \refitem { Jakobsen {\it et al.} 1991, {\it Ap.J.Lett.}, {\bf 369}, L63.}

  \refitem { Jeffery, D. J., Branch, D. 1990, in {\it Jerusalem Winter School
for Theoretical Physics, Vol 6., Supernovae,} ed. J.C. Wheeler, T. Piran,
and S. Weinberg, (Singapore: World Scientific) p. 149.}

  \refitem { Jeffery, D. J. 1989, {\it Ap.J.Supp}, {\bf 71}, 951.}

  \refitem { Johnson, H. L. 1965, {\it Ap.J.}, {\bf 141}, 923.}

  \refitem { Karp, H. H., Lasher, G., Chan, K. L., Salpeter, E. E. 1977,
{\it Ap.J.}, {\bf 214}, 161.}

 \refitem { Kharadze, E. K., Pskovsky, Y. P. 1989, {\it I.A.U. Circular 4802.}}

  \refitem { Kidger, M. 1988, {\it I.A.U. Circulars 4541 and 4613}.}

  \refitem { Kirshner, R. P. 1985, in {\it Lecture Notes in Physics:
Supernovae as Distance Indicators}, ed. N. Bartel (Berlin: Springer-Verlag),
 p. 171.}

  \refitem { Kirshner, R.P. 1990, in {\it Supernovae}, ed. A.G. Petschek (New
York: Springer-Verlag) p. 59.

\refitem { Kirshner, R. P., Arp, H. C., Dunlap, J. R. 1976, {\it Ap.J.},
 {\bf 207}, 44.}

  \refitem { Kirshner, R. P., Kwan, J. 1974, {\it Ap.J.}, {\bf 193}, 27.}

  \refitem { Kirshner, R. P., Kwan, J. 1975, {\it Ap.J.}, {\bf 197}, 415.}

  \refitem { Kraan-Korteweg, R. C. 1986, {\it Astr.Ap.Supp.}, {\bf 66}, 255.}

  \refitem { Leavitt, H. 1912, {\it H.C.O. Circular No. 173}.}

  \refitem { Leibundgut, B. 1991, in {\it Supernovae, Proceedings of the
 Tenth Santa Cruz Summer Workshop.}, ed. S. E. Woosley, (New York:
Springer-Verlag 1991) p. 751.}

 \refitem { Leibundgut, B., Kirshner, R. P., Filippenko, A. V., Shields,
J. C., Foltz, C. B., Phillips, M. M., Sonneborn, G. 1991, {\it Ap.J.Lett.},
 {\bf 371}, L23}

 \refitem { Leibundgut, B., Kirshner, R. P., Pinto, P. A., Rupen, M. P.,
Smith, R. C., Gunn, J. E., Schneider, D. P. 1991, {\it Ap.J.}, {\bf 372}, 521.}

  \refitem { Lynden-Bell, D., Faber, S. M., Burstein, D., Davies, R. L.,
 Dressler, A., Terlevich, R. J., Wegner, G. 1988, {\it Ap.J.}, {\bf 326}, 19.}

  \refitem { Mihalas, D. 1978. {\it Stellar Atmospheres}, (2nd ed., San
Francisco: Freeman).}

  \refitem { Miller, D. L., Branch, D. 1990, {\it A.J.}, {\bf 100}, 530.}

  \refitem { Oke, B., Gunn, J. 1983, {\it Ap.J}, {\bf 266},713.}

  \refitem { Panagia N., Gilmozzi, R., Macchetto, F., Adorf, H. M., Kirshner,
 R. P. 1991, {\it Ap.J.Lett.}, {\bf 380}, L23.}

  \refitem { Panagia, N. 1980, {\it I.A.U. Circular 3537}.}

  \refitem { Panagia, N. {\it et al.}  1980, {\it M.N.R.A.S.}, {\bf 192}, 861.}

  \refitem { Papaliolios, C., Karovska, M., Koechlin, L., Nisenson, P.,
Standley, C., Heathcote, S. 1989, {\it Nature}, {\bf 338}, 565.}

  \refitem { Pennypacker, C., Perlmutter, S. 1990, {\it I.A.U. Circular 4965.}}

  \refitem { Penston, M. V., Blades, J. C. 1980, {\it M.N.R.A.S.}, {\bf 190},
 51p.}

  \refitem { Pettini, M. {\it et al.} 1982, {\it M.N.R.A.S.}, {\bf 199}, 409.}

  \refitem { Phillips, M. M., Heathcote, S. R., Hamuy, M., Navarrete, M. 1988,
 {\it A.J.}, {\bf 95}, 1087.}

  \refitem { Phillips, M. M., Wells, L. A., Suntzeff, N. B., Hamuy, M.,
 Leibundgut, B., Kirshner, R. P., Foltz, C. B. 1992 {\it A.J. in press}}

  \refitem { Pierce, M. J., Tully, R. B. 1988, {\it Ap.J.}, {\bf 330}, 579.}

  \refitem { Reid, I. N., Strugnell, P. R. 1986, {\it M.N.R.A.S.}, {\bf 221},
887.}

\refitem { Ruiz-Lapuente, P., Kidger, M., Gomez, G., Canal, R., Lopez, M.
  1991, {\it Ap.J.Lett.}, {\bf 378}, L41.}

  \refitem { Ruiz-Lapuente, P. Kidger, M., Lopez, R., Canal, R. 1990,
{\it Astr.Ap.}, {\bf 100}, 782.}

  \refitem { Rybicki, G. B., Lightman, A. P. 1979. {\it Radiative Processes
in Astrophysics}, (New York: John Wiley and Sons).

  \refitem { Sadler, E. M., Simkin, S. M. 1988, {\it I.A.U. Circular 4563}.}

  \refitem { Sakurai, A. 1960, {\it Comm. in Pure and App. Math}, {\bf 13},
 353.}

\refitem { Sandage, A., Tammann, G. A. 1984, {\it Nature}, {\bf 307}, 326.}

  \refitem { Sandage, A., Tammann, G. A. 1990, {\it Ap.J.}, {\bf 365}, 1.}

  \refitem { Schechter, P. L. 1980, {\it A.J.}, {\bf 85}, 801.}

  \refitem { Schmutz, W., Abbot, D. C., Russell, R. S., Hamann, W. R.,
 Wessolowski, U. 1990, {\it Ap.J.}, {\bf 355}, 255.}

  \refitem { Schurmann, S. R., Arnett, W. D., Falk, S. W. 1979, {\it Ap.J.},
{\bf 230}, 11.}

  \refitem { Searle, L. 1971, {\it Ap.J.}, {\bf 168}, 327 }

  \refitem { Selby, E. A., McClatchey, R. M. 1972, {\it Atmospheric
Transmittance from .25 to 28.5 Micrometers Computer Code Lowtran 2} (Air
Force Cambridge Research Lab. Rept. AFCRL-72-0745).}

  \refitem { Shaviv, G., Wehrse, R., Wagoner, R. V. 1985, {\it Ap.J.},
{\bf 289}, 198.}

  \refitem { Sramek, D., van der Hulst, J. M., Weiler, K. W. 1980, {\it
I.A.U. Circular 3557}.}

  \refitem { Steidel, C. C., Rich, R. M., McCarthy, J. K. 1990, {\it A.J.},
{\bf 99}, 1476.}

 \refitem { Tammann, G. A., Leibundgut, B. 1990, {\it Astr.Ap.}, {\bf 236}, 9.}

  \refitem { Taylor, B. 1986, {\it Ap.J.Supp.}, {\bf 60}, 577.}

  \refitem { Thompson, L. A. 1982, {\it Ap.J.Lett.}, {\bf 257}, L63.}

  \refitem { Tonry, J. L. 1991, {\it Ap.J.Lett.}, {\bf 373}, L1.}

  \refitem { Tonry, J. L., Schechter, P. L. 1990, {\it A.J.}, {\bf 100}, 1794.}

  \refitem { Tonry, J. L., Ajhar, E.A., Luppino, G. A. 1990, {\it A.J.}, {\bf
100}, 1416.}

  \refitem { Tonry, J. L., Schneider, D. P. 1988, {\it Ap.J.}, {\bf 96}, 807.}

  \refitem { Tully, R. B., Fisher, J. R. 1977, {\it Astr.Ap.}, {\bf 54}, 661.}

  \refitem { Uomoto, A., Kirshner, R. P. 1986, {\it Ap.J.}, {\bf 388}, 685.  }

  \refitem { Wagoner, R. V. 1981, {\it Ap.J.Lett.}, {\bf 250}, L65.}

  \refitem { Wagoner, R. V. 1991, in {\it Supernovae, Proceedings of the Tenth
 Santa Cruz Summer Workshop.}, ed. S. E. Woosley, (New York: Springer-Verlag)
 p 741.}

  \refitem { Walker, A. R. 1987, {\it M.N.R.A.S.}, {\bf 225}, 627.}

  \refitem { Walker, A. R., Mack, P. 1988, {\it A.J.}, {\bf 96}, 1362.}

  \refitem { Wamsteker, W. 1972, {\it Astr.Ap.}, {\bf 19}, 99.}

  \refitem { Welch, D. L., McLaren, R. A., Madore, B. F., McAlary, C. W.
1987, {\it Ap.J.}, {\bf 321}, 162.}

  \refitem { Whitford, A. E.  1958. {\it A.J.}, {\bf 63}, 201.}

  \refitem { Winzer, J. E. 1974, {\it R.A.S.C.}, {\bf 68}, 36.}

  \refitem { Wood, R., Andrews, P. J. 1974, {\it M.N.R.A.S.}, {\bf 167}, 13.}

  \refitem { Woosley, S. E. 1988, {\it Ap.J.}, {\bf 330}, 218.}

  \refitem { de Vaucouleurs, G., Beardsley, B., Buta, R., Smith, B. 1979, {\it
I.A.U. Circular 3361}.}

\vfill
\eject

\nopagenumbers
\centerline {\bf Figure Captions}
\vskip .1 in

\noindent {\bf Figure 1:} Comparison of free-free, bound-free and electron
scattering
opacity versus wavelength from a model of a 15 $M_\odot$ red supergiant
10 days after explosion.
\vskip .15 in

\noindent {\bf Figure 2:}  The time variation of distance correction factors
derived from fitting optical and infrared photometry to models of SN 1987A, and
15$M_\odot$ and 25$M_\odot$ red supergiant supernovae.
\vskip .15 in

\noindent {\bf Figure 3:}  Distance correction factors, $\zeta_{emp}$,
empirically derived from SN 1987A, shown as a function of the optical
color temperature.

\vskip .15 in

\noindent {\bf Figure 4:}  Optical distance correction factors plotted as a
function of
color temperature derived empirically, and from models of SN 1987A, and
15$M_\odot$
and 25$M_\odot$ red supergiant supernovae.  The line of best fit is also
plotted.

\vskip .15 in

\noindent {\bf Figure 5:}  $(B-V)$ color evolution of five SN II-P.
The unreddened supernovae; SN 1968L, SN 1969L, and SN 1990ae
all show a uniform evolution. SN 1973R and SN 1990E are reddened, and
show a $(B-V)$ color excess.
\vskip .15 in

\noindent {\bf Figure 6a:}  The percent change in distance as a function of
visual extinction for SN 1968L, SN 1969L, SN 1970G, and SN 1979C.
\vskip .15 in

\noindent {\bf Figure 6b:}  The percent change in distance as a function of
visual extinction for SN 1973R, SN 1980K, SN 1990E, and SN 1990ae.
\vskip .15 in

\noindent {\bf Figure 7:}  Comparison of the temperature of the supernova
continuum
as derived from the $(B-V)$ color and optical spectrum.  The dotted line is
$T_{B-V}=T_{Spec}$.
\vskip .15 in

\noindent {\bf Figure 8a:}  Distance as a function of time for
SN 1968L, SN 1969L, SN 1970G, and SN 1988A.
\vskip .15 in

\noindent {\bf Figure 8b:}  Distance as a function of time for
SN 1973R, SN 1979C, SN 1980K, and SN 1990E.
\vskip .15 in

\noindent {\bf Figure 9:}  Distance as a function of time for
SN 1987A using infrared and optical photometry.

\vskip .15 in

\noindent {\bf Figure 10:} A 10 minute exposure in $R$ at the KPNO 2.1 meter
telescope showing SN 1990ae in its anonymous parent galaxy on 1990 October 23.
 \vskip .15 in

\noindent {\bf Figure 11:}  Spectrum of SN 1990ae from the
MMT Blue Channel on 1990 October 19,  and MMT Red Channel on 1990 November 12.

\vskip .15 in

\noindent {\bf Figure 12:}  The Virgo corrected velocities for 8 galaxies with
SN II (SN 1990ae is excluded) are plotted against their distances derived using
the Expanding Photosphere Method.
Lines for Hubble constants ranging from 50 km sec$^{-1}$ Mpc$^{-1}$ to 80 km
sec$^{-1}$ Mpc$^{-1}$ are shown.
\vskip .15 in

\noindent {\bf Figure 13:}   The Virgo corrected velocities for 9 galaxies with
SN II are plotted
against their distances derived using the Expanding Photosphere Method.
Lines for Hubble constants ranging from 50 km sec$^{-1}$ Mpc$^{-1}$ to 80 km
sec$^{-1}$ Mpc$^{-1}$ are shown.

\end{document}